# Evolution of Raman G and G' (2D) Modes in Folded Graphene Layers

Chunxiao Cong,[1] Ting Yu[1,2,3] [†]

[1]*Division of Physics and Applied Physics, School of Physical and Mathematical Sciences, Nanyang Technological University, 637371, Singapore;*

[2]*Department of Physics, Faculty of Science, National University of Singapore, 117542, Singapore;*

[3]*Graphene Research Center, National University of Singapore, 117546, Singapore*

[†]Corresponding authors. E-mail: yuting@ntu.edu.sg

**ABSTRACT**

Bernal- and non-Bernal-stacked graphene layers have been systematically studied by Raman imaging and spectroscopy. Two dominant Raman modes, G and G' (or 2D) of folded graphene layers exhibit three types of spectral features when interlayer lattice mismatches, defined by a rotational angle varies. Among these folded graphene layers, the most interesting one is the folded graphene layers that present an extremely strong G mode enhanced by a twist-induced Van Hove singularity. The evolution of Raman G and G' modes of such folded graphene layers are probed by changing the excitation photon energies. For the first time, doublet splitting of the G' mode in folded double-layer (1 + 1) and of the G mode in folded tetra-layer (2 + 2) graphene are clearly observed and discussed. The G' mode splitting in folded double-layer graphene is attributed to the coexistence of inner and outer scattering processes and the trigonal warping effect as well as further downwards bending of the inner dispersion branch at visible excitation energy. While the two peaks of the G mode in folded tetra-layer graphene are assigned to Raman-active mode ($E_{2g}$) and lattice mismatch activated infrared-active mode ($E_{1u}$), which is further verified by the temperature-dependent Raman measurements. Our study provides a summary and thorough understanding of Raman spectra of Bernal- and non-Bernal-stacked graphene layers and further demonstrates the versatility of Raman spectroscopy for exploiting electronic band structures of graphene layers.

# I. INTRODUCTION

Electronic band structures of graphene layers are remarkably influenced by the ways of these carbon atomic layers stacking themselves. The most stable and common one is AB- (or Bernal) stacking. From pristine monolayer to Bernal-stacked bilayer and few-layer graphene, electronic band structures show significant differences and can be effectively probed by investigating their Raman spectral features such as relative intensities, linewidths, line shapes and peak positions of Raman G and G' (or 2D) modes through the strong electron-phonon coupling.[1-6] Such unique optical response promises Raman spectroscopy to be a widely adapted technique to quickly and precisely identify thickness of pristine Bernal-stacked graphene layers. However, when such graphene layers are subjected to some local electrical or mechanical perturbations, for example, varying local electrical potential by coating molecules or applying electrical gate,[7-13] or expanding the substrate where graphene layers anchoring by applying uniaxial strain,[14-17] Raman modes of G and G' could change remarkably. Thus, one should pay special attention to determine number of layers of graphene by using the spectral features of Raman G and G' modes. In addition to the perfect Bernal stacking, graphene layers may naturally or artificially stack themselves into other sequences, which also lead to the different Raman spectral features compared to Bernal-stacked ones. The two most well-known non-Bernal stacked graphene layers are ABC-stacked trilayer graphene and 1+1 folded or twisted double-layer graphene (fDLG or tDLG). As being widely used in the studies of graphite and carbon nanotubes,[18-19] Raman spectroscopy once again has demonstrated its special and powerful ability to probe the electronic band structures of such two interesting two-dimensional carbon systems. Quick and accurate identification and even visualization of ABC-stacked trilayer graphene domains from Bernal-stacked ones by Raman imaging and spectroscopy have been successfully demonstrated.[20-21] Modulation of electronic band structures by a stacking defect such as a twist was predicated by the theoretical study at the early stage of graphene investigation[22] and the proposed remaining of linear dispersion and reduction of Fermi velocity were evidenced by the Raman spectroscopy study of the G' mode in a 1+1 fDLG soon afterwards.[23] Later, extremely strong G mode, *i.e.* tens of times stronger than G' mode, was observed in the fDLG.[24] On the contrary of the well-known dispersive D mode, a non-dispersive defect (rotational stacking) mode was seen and creatively explained as the rotational angle dependent wave vector assisted double resonant scattering process.[25] To prepare fDLG of more folding or rotational angles, a flipping over technique was developed by using an atomic force microscope (AFM).[26-27] New peaks, next to the G mode at both lower and higher frequency regions appeared in such fDLG, which are also activated by the period static potential defined by the rotational angles or the wavevector in the supperlattice.[26-27] Large-scale growth of graphene layers by chemical vapor deposition (CVD) offers feasibility of producing tDLG with a wide range of twisting angles by transferring one layer followed by the other or locating the as-grown tDLG with multiple domains.[28-30] More direct evidence of the change of electronic band structure by rotational stacking such as the twist-induced Van Hove singularities (VHS) was given by a scanning tunneling microscopy (STM)/spectroscopy (STS) study of twisted CVD graphene double layers.[31] From transmission electron microscopy (TEM) diffraction patterns, orientations of domains and consequently twisting angles between two graphene layers can be precisely determined. Combining TEM and Raman spectroscopy, the Raman G and G' modes of tDLG with twisting angels from 0 to 30 degrees were systematically studied.[28,

[30] The twist-induced VHS could be clearly reflected by the dramatic enhancement of the Raman G mode.[30] Meanwhile, the spectral features such as linewidths, frequencies and intensities also show dependence on the twisting angles, demonstrating that the Raman G and G' modes can be used to identify the twisting angles.[28, 30] Most recently, CVD grown tDLG consisting of single domain in each layer were studied by Raman spectroscopy.[32-33] Twist-induced Raman modes of low and intermediate vibrational frequencies were also observed.[29, 32-33]

In this work, we focus on the dominant Raman G and G' modes and exploit their evolution in graphene layers of different stacking orders including both Bernal and non-Bernal ones. In addition to 1+1 folded double-layer graphene, we also studied 2+2 folded tetra-layer graphene (f4LG). Here the "2" refers to Bernal-stacked bilayer graphene. Such 2+2 f4LG are hardly prepared by CVD and rarely studied. By carefully checking the spectral features such as relative intensities, line shapes, linewidths and peak positions, three patterns of Raman G and G' modes are identified, corresponding to the graphene layers of three types of rotational stacking defined by three ranges of rotational angles, $\theta$. In detail, for the excitation photon energy of 2.33 eV, the first group has a relatively small rotational angle below 4 degrees (named as $\theta_{small}$), the second group possesses a rotation angle of around 11 degrees (named as $\theta_{medium}$), and rotational angles are more than 20 degrees for the third group (named as $\theta_{large}$). Among these different types of folded graphene layers, the most interesting one is the ones showing strong enhancement of the G mode intensity due to the twist induced VHS (our labeling, $\theta_{medium}$). Doublet splitting of the G' mode for 1+1 fDLG $\theta_{medium}$ samples and of the G mode for 2+2 f4LG $\theta_{medium}$ samples are noticed through the investigation of the evolution of these two modes under different excitation photon energies. Our further polarization- and temperature-dependent Raman spectroscopy studies reveal that the splitting is due to the coexistence of inner and outer scattering processes and the trigonal warping effect for the G' mode and Raman-active mode ($E_{2g}$) and stacking defect-activated infrared (IR)-active mode ($E_{1u}$) for the G mode.

## II. EXPERIMENT

All the graphene layers in this work were prepared by the mechanical cleavage of graphite and transferred onto a 300 nm SiO$_2$/Si substrate. The folded graphene layers were self-formed by accident during the mechanical exfoliation process. An optical microscope was used to locate the folded thin layers, and the number of layers of the unfolded part was further identified by white light contrast spectra and Raman spectroscopy.[34] The Raman images were acquired using a WITec CRM200 Raman system with a 600 lines/mm grating and a piezocrystal controlled scanning stage under 532 nm ($E_{laser}$ = 2.33 eV) laser excitation. A grating of 2400 lines/mm was used for single Raman spectrum measurement under different excitation energies to achieve high spectral resolution. For the room temperature Raman measurements, an objective lens of ×100 magnification and 0.95 numerical aperture (NA) was used, and the laser spot is of ~500 nm in diameter. The laser power was kept below 0.1 mW on the sample surface to avoid laser-induced heating. For the temperature dependent Raman measurement, a long-working-distance ×50 objective with NA of 0.55 is used and the sample temperature was controlled by a programmable hot stage HFS600E from Linkam Scientific Instruments.

## III. EXPERIMENTAL RESULTS AND DISCUSSION

Though the variation of the general Raman spectral features of the G and G' modes were observed and discussed in tDLG grown by CVD over a wide range of rotational angles, more details are subjected to be further probed.[30, 35] Figure 1 presents the Raman images and spectra of three types of fDLG and pristine single layer graphene (SLG) obtained under the excitation photon energy of 2.33 eV together with the optical image and the schematic illustration. In this work, unless specially clarified, all Raman images were plotted by extracting the spectral features through a single Lorentzian line shape fitting. The folding angles were measured from both the optical and AFM images as previously reported[24] (more details in support information and Fig. S1) and are shown next to the corresponding Raman spectra (see Fig. 1c). Comparing to the SLG, all G' modes of fDLG show a blue-shift due to reduction of Fermi velocity.[22-23] For the sample shown in panel *a* and labeled as $\theta_{small}$, the Raman D mode image (a5) further indicates the dominant edge orientations as also illustrated schematically by light blue honeycomb in (a5).[36-38] As shown, for the fDLG of a small twisting angle (less than 4 degrees), our $\theta_{small}$, the integrated intensity of the G' mode is even weaker than that of SLG (Fig. 1a6) whereas the linewidth is much larger than that of SLG (Fig. 1a8). The corresponding Raman spectrum (Fig. 1c) of the 1+1 fDLG $\theta_{small}$ sample displays a broad asymmetric G' peak consisting of multiple sub-peaks, instead of a narrow single symmetric peak appearing in the spectrum of SLG. This might be due to the formation of a Bernal-stacked bilayer graphene (BLG)-like electronic structure led by the strong interlayer interaction and highly overlapping of two cones of each layer.[39] More details of the asymmetric Raman G' mode of 1+1 $\theta_{small}$ fDLG under different excitation energies are shown in the support information together with that of Bernal-stacked BLG (see Fig. S2). It shows that the broad and asymmetric Raman G' mode of 1+1 $\theta_{small}$ fDLG can be fitted well by four Lorentzian peaks, which is usually used for fitting of the Raman G' mode of AB-stacked BLG. And the frequencies of all of the four sub-peaks of 1+1 $\theta_{small}$ fDLG increase linearly with increase of the excitation energy, which is similar to that of AB-stacked BLG. This is in some ways to indicate the similarity of the electronic structures between the 1+1 $\theta_{small}$ fDLG and the AB-stacked BLG. The sample shown in panel *b* contains two folded areas as also illustrated in Fig. 1b5. The most obvious difference of the Raman spectral features of these two fDLG is the remarkable contrast of G mode intensities (Fig. 1b2) or the relative intensities between the G and G' mode (Fig. 1c). The significant enhancement of the G mode in our 1+1 fDLG $\theta_{medium}$ sample can be well interpreted by the resonant effect between the conduction and valence twist-induced VHS.[30] A small $R$ peak can be also seen in this fDLG $\theta_{medium}$ sample, which is activated and defined by the twist-induced wavevector in the supperlattice.[26] Very practically meaningful, this $R$ peak could be a good indicator of rotational angles as demonstrated previously.[27, 29] The interlayer coupling in our fDLG $\theta_{large}$ samples is relative weak. The G' mode remains the same line shape as that of SLG and is much stronger than the G mode. It should be noticed that the types of our clarifications and Raman spectral patterns are both rotational-angles- and excitation-photon-energies-dependent. For different laser lines, the same folded type may appears at different rotational angles.

To further exploit the Raman G and G' modes of these three types of fDLG, the evolutions of these two modes under different excitation energies are studied. Figure 2 shows the laser-excitation-energy-dependent Raman spectra of fDLG together with that of SLG as a comparison. Owing to the dispersive nature and the double resonant scattering process, all G' modes show blue-shift as the increase of the excitation photon energies while the zone center G phonons keep a constant vibrational frequency. Our focus of this part is the $\theta_{medium}$ sample as it expresses the most remarkable change (Fig. 2c). Firstly, the small $R$ peak does not shift when changing the excitation energies. As reported, this is because that the $R$ peak is mediated by the twist-induced wavevector, which is strictly defined by the rotational angles.[26] So, once the rotational angle is fixed, the wavevector is defined, and then the momentum and subsequently the energy of phonons are fixed or selected, leading to the non-dispersion of the $R$ peak.[26] Using this criterion and the plots in Ref. 26, the position of 1506 cm$^{-1}$ corresponds to a rotational angle of 11 degrees which is very close to our measured value of 10.8 degrees. Very different to $\theta_{small}$ (BLG-like) or $\theta_{large}$ (SLG-like), the G' mode of $\theta_{medium}$ fDLG samples present a very obvious doublet splitting under a higher excitation energy, 2.54 eV in this work. The doublet splitting of G' mode has been observed in suspended[40-41] or uniaxially stretched SLG.[42-44] In the latter case, the G' peaks would show substantial red-shift and broadening,[15-17] which are absent in our data. Therefore, we would temporarily consider our observations along the track of the arguments on bimodal G' mode line shape in suspended single layer graphene. By suspending the single layer graphene and consequently suppressing the substrate induced unintentional doping, the doping induced broadening effects on Raman G' modes could be weakened and the intrinsic narrow sub-peaks will be resolved if there is any.[40-41] If the excitation photon energy is high enough to pump electrons to a level where the Dirac cone is distorted and a triangular shaped equal energy contours present, the so-called electronic trigonal warping effect together with phononic trigonal warping effect will play critical roles in the electron-phonon double resonant scattering process and could be responsible for the G' mode splitting.[40-41,45]

To further understand this doublet splitting of the G' mode in the fDLG $\theta_{medium}$ sample, we performed careful curve fitting of the spectra under different excitation energies and polarization-dependent Raman spectroscopy measurements. From the electronic band structure of pristine SLG, the trigonal warping effect becomes more obvious at a relatively higher energy level and even the electronic dispersions are no longer perfectly linear, for example bending towards lower energy level along the *K-M* direction, corresponding to the inner scattering process whereas the dispersion along the *K-Γ* direction for the outer scattering process bending upwards.[45] Such evolution of electronic band structure at relatively high energy level should immediately cause the higher frequency G' peak (G'+) moving further away from the lower frequency one (G'-).[40, 45] It should be noticed that the phonon trigonal warping effect could also affect the positions of two G' peaks from the inner and outer scattering processes. The previous study has clearly demonstrated that the separation of these two peaks still increases when the excitation energy is beyond 2.3 eV by considering both phononic and electronic trigonal warping effects.[45] Our previous experimental study on the suspended SLG clearly evidenced this, in which the separation of the two G' peaks increases from 12 cm$^{-1}$ in the visible excitation photon energy to ~20 cm$^{-1}$ at ultraviolet light of 3.49 eV.[40] As shown in Figure 3a, with the increase of the excitation energies, the Raman frequency separation between the G'+ mode and the G'- mode of $\theta_{medium}$ fDLG indeed increases. This agrees well with the theoretical predication of SLG at higher excitation energies.[40, 45] Slightly different to the previous studies,[40-41] the increment of the frequency difference in the visible excitation photon energy range for our $\theta_{medium}$ fDLG sample could possibly result from the further downwards bending of the dispersion along the *K-M* direction induced by the twist even in the visible excitation photon energy range besides the phononic and electronic trigonal warping effects.[39] Therefore, we tentatively assign the higher frequency component (G'+) to the inner scattering process and the lower frequency component (G'-) to the outer scattering process. The relative intensity of the G'+ feature (inner process) over the G'- feature (outer process) is considerably complicated in the folded or twisted graphene layers because there is anisotropic contribution to the G' mode in the phonon Brillouin zone and such nonuniformity varies when the rotational angles change.[35] This is out of the scope of this work. The schematic diagrams of the inner and outer scattering processes of the G' mode in pristine SLG and our $\theta_{medium}$ fDLG are shown in Fig. S3. Another consequence of the trigonal warping effects, also being an effective way to probe such effect is the different response of G'- and G'+ modes to the polarization configuration of the incident and scattering lights. Due to the triangular shape or the different curvatures of the electronic band structures facing to the *K-M* (inner scattering) and *K-Γ* (outer scattering) directions, the anisotropic optical absorption around the *K* point,[46] and also the dependence of Raman G' intensity on involved phonon wavevector directions,[45] the relative intensity of outer scattering process over inner scattering process could be very different at different polarization conditions. As demonstrated in Fig. 3b and c, under the parallel polarization (XX), the G'+ is more dominant, which perfectly agree with the previous findings of Raman G' band of SLG at higher excitation energies.[45] Considering the above discussion about the dependence of the G' mode of the $\theta_{medium}$ fDLG on the excitation energy and polarization, we tentatively attribute the doublet splitting of the G' band here to the coexistence of the inner and the outer scattering processes and the trigonal warping effects as well as further downwards bending of the inner dispersion branch at visible excitation energy.[39] The arguments of outer and inner scattering processes in the G' phonons have been a long-lived debate. Our findings demonstrate that the folded or twisted graphene layers could be an interesting system for such topic. More systematic study is needed.

Though 1+1 fDLG or tDLG has been intensively studied, 2+2 f4LG is rarely probed. As shown in Figure 4, very similar to the fDLG, 2+2 f4LG also exhibits three types. For the 2+2 $\theta_{medium}$ f4LG, our simulation clearly reveals the formation of VHS (to be discussed somewhere else).[47] Different to the fDLG, the G mode of $\theta_{medium}$ f4LG shows a doublet splitting and two sub-peaks are clearly resolved (Fig. 4c). It is known that G mode splitting or two-peak G mode occurs when the graphene layers are highly and asymmetrically doped at top and bottom layers, which causes the breaking inverse symmetry and consequently the phonon mixing.[48-51] Applying substantially large uniaxial strain could also split the G mode by breaking the symmetry of the lattice and subsequently breaking the degeneracy of the two-fold symmetric $E_{2g}$ phonons.[15-16, 48] However, for this work, neither of the above two substances should apply since we did not highly dope or substantially stretch the samples as reflected by the absence of large amount of blue-shift, the response of doping[12] nor the red-shift of the G mode, the response of tensile strain.[14, 17] Two Raman modes, D' and R' also locate near the G mode and at high frequency side. However, the truth that both two peaks are much weaker than that of G peak[26, 29] unambiguously indicates that neither of them could be responsible for the observed very sharp and strong G+ mode here.

To further unveil the origin of the doublet splitting of the G mode in our $\theta_{medium}$ f4LG, firstly, we conducted excitation-energy-dependent Raman measurements and plotted the Raman spectra in Fig. 5a-c. It is clear enough that the second G peak (G+) appears when the normal G mode (G-) is resonantly enhanced and simultaneously the weak $R$ peak is also activated. It is noticed that the position of $R$ peak here (1464 cm$^{-1}$) is smaller than that of the $R$ peak in our $\theta_{medium}$ fDLG, meaning a larger rotational angle should be expected,[29] which is clearly demonstrated by our measurements (Fig. 1c and Fig. 4c). Furthermore, the same as the normal G mode, this new peak (G+) is also non-dispersive, indicating it might be a zone center phonon too. The linewidths and the intensity ratios of two split G peaks could be sensitive indicators of the asymmetric doping levels, where phonon mixing may exist.[48] The detailed curve fitting (Fig. 5f) of the G mode in our $\theta_{medium}$ f4LG provides more spectral parameters of the split G peaks: the ratio of full width at half maximum (FWHM) between G- peak and G+ peak is ~2.3 and the relative intensity ratio of G+ peak to G- peak ($I_{G+}/I_{G-}$) is around 0.3. Comparing these features together with the parameters of another $\theta_{medium}$ 2+2 f4LG (Fig. S4 and Table S1) with the spectral parameters of the asymmetrically doped BLG, it is noticed that our case is very much different to the optical phonon mixing in asymmetrically doped bilayer graphene.

For example, the linewidth ratios (FWHM$_{G-}$/FWHM$_{G+}$) of our two $\theta_{medium}$ 2+2 f4LG samples are around 2.3, but their I$_{G+}$/I$_{G-}$ could be very different (1.25 in Fig. S4 and 0.31 in Fig. 5f). On a clear contrast, for such ratio of G- width over G+ width (~2.3), the I$_{G+}$/I$_{G-}$ of two G peaks induced by the phonon mixing is always very small, like near 0.1.[48, 50] The line shapes of the Raman G band are also very different between our $\theta_{medium}$ 2+2 f4LG and the one caused by the phonon mixing in the asymmetrically doped BLG. For example, for phonon mixing, when the peak widths are quite different, one of the split peaks is so dominant that the overall spectra appear like one broad peak [48, 50] while one narrow and one broad peaks obviously present in the spectra of our samples. The different relative intensities among our folded graphene samples might be dependent on the twisting angles and consequently the evolution of the bandstructures for a given excitation energy, which will be further studied through our ongoing projects.

The observation of the narrow G+ mode rules out the possibilities that this G+ peak is due to strain as the G+ peak appears as broad as G- peak for the strain case.[15-16] Considering the above findings and discussions, we speculate the G+ mode in the $\theta_{medium}$ f4LG is an IR-active mode $E_{1u}$, which might be activated and enhanced by the stacking defect, the twist of a unique twisting angle, *i.e.* $\theta_{medium}$ in this work. Our temperature-dependent Raman measurements strongly support this speculation. The temperature-dependent Raman spectra in the G mode region with fitted curves of $\theta_{medium}$ f4LG are shown in Fig. S5. Figures 5d and e present the FWHM and the peak position separation of the G- ($E_{2g}$) mode and the G+ ($E_{1u}$) mode, respectively, as a function of temperatures. As can be clearly seen, with the decrease of the temperatures, the linewidth of G+ mode decreases while the G- mode becomes slightly broader. While the vibrational energy separation of these two modes increases as the consequence of a bit faster hardening of G+ mode comparing to G- mode with the decrease of the temperatures. This shows a good agreement with the previous studies, and could be explained by the different electron-phonon anharmonic scattering in the Raman- and IR-active modes for the linewidths, and the presence (absence) of the coupling between $E_{1u}$ ($E_{2g}$) phonon and a low wavenumber out of plane optical phonon for the separation of peak positions.[52] Though in this work, we demonstrate the enhancement of G mode and the splitting of G' mode in a 1+1 fDLG of rotational angle of 10.8 degrees and the splitting of G mode in a 2+2 f4LG of rotational angle of 12.4 degrees, it is worth to note that the rotational angles, $\theta_{medium}$, which initiate the enhancement of the G mode and the doublet splitting of G' and G modes in folded graphene layers are dependent on the excitation photon energies and may vary by a few degrees under different excitation energies.

## IV. CONCLUSIONS

In summary, we have systematically investigated the Raman G and G' modes of 1+1 and 2+2 folded graphene layers. Three types of folded samples corresponding to three ranges (small, medium and large) of rotational angles are classified by evaluating and comparing their Raman spectral features to those of pristine SLG and Bernal-stacked bilayer graphene. The evolution of the Raman G and G' modes in fDLG and f4LG under different excitation energies are probed. A doublet splitting of the G' mode in the $\theta_{medium}$ fDLG and the G mode in the $\theta_{medium}$ f4LG is observed and well explained by the co-existence of (i) the inner and the outer scattering modes and the trigonal warping effects as well as further downwards bending of the inner dispersion branch at visible excitation energy and (ii) the Raman-active and the stacking defect activated IR-active mode, respectively, through the systematic investigations of the polarization and temperature-dependent Raman spectroscopy. This work provides (i) the overall picture of the Raman spectra of the folded graphene layers, (ii) the evolution of the dominant Raman modes and (iii) the detailed understanding of the doublet splitting of the G and G' modes in the $\theta_{medium}$ folded graphene layers. It successfully demonstrates that Raman image/spectroscopy is indeed a unique and powerful tool for probing the electron-phonon coupling and electronic band structures for both Bernal- and non-Bernal-stacked graphene layers.

**Acknowledgements**

This work is supported by the Singapore National Research Foundation under NRF RF Award No. NRF-RF2010-07 and MOE Tier 2 MOE2012-T2-2-049. The authors gratefully thank the valuable help of Dr Jeil Jung for the density of states (DOS) simulation.

References

[1] A. C. Ferrari and D. M. Basko, Nat Nanotechnol **8**, 235 (2013).

[2] A. C. Ferrari *et al.*, Phys Rev Lett **97**, 187401 (2006).

[3] L. M. Malard, J. Nilsson, D. L. Mafra, D. C. Elias, J. C. Brant, F. Plentz, E. S. Alves, A. H. C. Neto, and M. A. Pimenta, Phys Status Solidi B **245**, 2060 (2008).

[4] M. S. Dresselhaus, A. Jorio, and R. Saito, Annu Rev Conden Ma P **1**, 89 (2010).

[5] L. M. Malard, M. A. Pimenta, G. Dresselhaus, and M. S. Dresselhaus, Phys Rep **473**, 51 (2009).

[6] A. Gupta, G. Chen, P. Joshi, S. Tadigadapa, and P. C. Eklund, Nano Lett **6**, 2667 (2006).

[7] X. C. Dong *et al.*, Phys Rev Lett **102**, 135501 (2009).

[8] X. C. Dong, D. L. Fu, W. J. Fang, Y. M. Shi, P. Chen, and L. J. Li, Small **5**, 1422 (2009).

[9] N. Peimyoo, T. Yu, J. Z. Shang, C. X. Cong, and H. P. Yang, Carbon **50**, 201 (2012).

[10] J. Yan, Y. B. Zhang, P. Kim, and A. Pinczuk, Phys Rev Lett **98**, 166802 (2007).

[11] J. Yan, E. A. Henriksen, P. Kim, and A. Pinczuk, Phys Rev Lett **101**, 136804 (2008).

[12] A. Das *et al.*, Nat Nanotechnol **3**, 210 (2008).

[13] S. Pisana, M. Lazzeri, C. Casiraghi, K. S. Novoselov, A. K. Geim, A. C. Ferrari, and F. Mauri, Nat Mater **6**, 198 (2007).

[14] T. Yu, Z. H. Ni, C. L. Du, Y. M. You, Y. Y. Wang, and Z. X. Shen, J Phys Chem C **112**, 12602 (2008).

[15] T. M. G. Mohiuddin *et al.*, Phys Rev B **79**, 205433 (2009).

[16] M. Y. Huang, H. G. Yan, C. Y. Chen, D. H. Song, T. F. Heinz, and J. Hone, P Natl Acad Sci USA **106**, 7304 (2009).

[17] Z. H. Ni, T. Yu, Y. H. Lu, Y. Y. Wang, Y. P. Feng, and Z. X. Shen, Acs Nano **2**, 2301 (2008).

[18] S. Reich and C. Thomsen, Philos T R Soc A **362**, 2271 (2004).

[19] A. G. Souza *et al.*, Phys Rev B **65**, 085417 (2002).

[20] C. H. Lui, Z. Q. Li, Z. Y. Chen, P. V. Klimov, L. E. Brus, and T. F. Heinz, Nano Lett **11**, 164 (2011).

[21] C. X. Cong, T. Yu, K. Sato, J. Z. Shang, R. Saito, G. F. Dresselhaus, and M. S. Dresselhaus, Acs Nano **5**, 8760 (2011).

[22] J. M. B. L. dos Santos, N. M. R. Peres, and A. H. Castro, Phys Rev Lett **99**, 256802 (2007).

[23] Z. H. Ni, Y. Y. Wang, T. Yu, Y. M. You, and Z. X. Shen, Phys Rev B **77**, 235403 (2008).

[24] Z. H. Ni, L. Liu, Y. Y. Wang, Z. Zheng, L. J. Li, T. Yu, and Z. X. Shen, Phys Rev B **80**, 125404 (2009).

[25] A. K. Gupta, Y. J. Tang, V. H. Crespi, and P. C. Eklund, Phys Rev B **82**, 241406 (2010).

[26] V. Carozo, C. M. Almeida, E. H. M. Ferreira, L. G. Cancado, C. A. Achete, and A. Jorio, Nano Lett **11**, 4527 (2011).

[27] V. Carozo *et al.*, Phys Rev B **88**, 085401 (2013).

[28] R. W. Havener, H. L. Zhuang, L. Brown, R. G. Hennig, and J. Park, Nano Lett **12**, 3162 (2012).

[29] J. Campos-Delgado, L. G. Cancado, C. A. Achete, A. Jorio, and J. P. Raskin, Nano Res **6**, 269 (2013).

[30] K. Kim *et al.*, Phys Rev Lett **108**, 246103 (2012).

[31] G. H. Li, A. Luican, J. M. B. L. dos Santos, A. H. C. Neto, A. Reina, J. Kong, and E. Y. Andrei, Nat Phys **6**, 109 (2010).

[32] R. He *et al.*, Nano Lett **13**, 3594 (2013).

[33] C. C. Lu, Y. C. Lin, Z. Liu, C. H. Yeh, K. Suenaga, and P. W. Chiu, Acs Nano **7**, 2587 (2013).

[34] Z. H. Ni, H. M. Wang, J. Kasim, H. M. Fan, T. Yu, Y. H. Wu, Y. P. Feng, and Z. X. Shen, Nano Lett **7**, 2758 (2007).

[35] Sinisa Coh, Liang Z. Tan, Steven G. Louie, and M. L. Cohen, Phys Rev B **88**, 165431 (2013).

[36] Y. M. You, Z. H. Ni, T. Yu, and Z. X. Shen, Appl Phys Lett **93**, 163112 (2008).

[37] C. X. Cong, T. Yu, and H. M. Wang, Acs Nano **4**, 3175 (2010).

[38] L. G. Cancado, M. A. Pimenta, B. R. A. Neves, M. S. S. Dantas, and A. Jorio, Phys Rev Lett **93**, 247401 (2004).

[39] E. S. Morell, J. D. Correa, P. Vargas, M. Pacheco, and Z. Barticevic, Phys Rev B **82**, 121407 (2010).

[40] Z. Q. Luo, C. X. Cong, J. Zhang, Q. H. Xiong, and T. Yu, Appl Phys Lett **100**, 243107 (2012).

[41] S. Berciaud, X. L. Li, H. Htoon, L. E. Brus, S. K. Doorn, and T. F. Heinz, Nano Lett **13**, 3517 (2013).

[42] D. Yoon, Y. W. Son, and H. Cheong, Phys Rev Lett **106**, 155502 (2011).

[43] O. Frank *et al.*, Acs Nano **5**, 2231 (2011).

[44] R. Narula, N. Bonini, N. Marzari, and S. Reich, Phys Rev B **85**, 115451 (2012).

[45] P. Venezuela, M. Lazzeri, and F. Mauri, Phys Rev B **84**, 035433 (2011).

[46] A. Gruneis, R. Saito, G. G. Samsonidze, T. Kimura, M. A. Pimenta, A. Jorio, A. G. Souza, G. Dresselhaus, and M. S. Dresselhaus, Phys Rev B **67**, 165402 (2003).

[47] C. X. cong and T. Yu, arXiv:1312.6928 (2013).

[48] W. J. Zhao, P. H. Tan, J. Zhang, and J. A. Liu, Phys Rev B **82**, 245423 (2010).

[49] T. Ando and M. Koshino, J Phys Soc Jpn **78**, 034709 (2009).

[50] J. Yan, T. Villarson, E. A. Henriksen, P. Kim, and A. Pinczuk, Phys Rev B **80**, 241417 (2009).

[51] P. Gava, M. Lazzeri, A. M. Saitta, and F. Mauri, Phys Rev B **80**, 155422 (2009).

[52] P. Giura, N. Bonini, G. Creff, J. B. Brubach, P. Roy, and M. Lazzeri, Phys Rev B **86**, 121404 (2012).

**Figure captions**

**Figure 1.** (a1) and (b1) show the optical images of different folded double-layer graphene sheets. Panels a2(a6), a3(a7) and a4(a8) are Raman images of the G(G') mode integrated intensity, the G(G') mode frequency, and the G(G') mode width of the corresponding folded double-layer graphene samples shown in panel a1. Panels b2(b6), b3(b7) and b4(b8) are Raman images of the G(G') mode intensity, the G(G') mode frequency, and the G(G') mode width of the corresponding folded double-layer graphene samples shown in panel b1. (a5) is Raman image of the D mode intensity of the edges of the single layer sheet shown in (a1), which indicates the crystal orientation (illustrated by the light blue honeycomb) of the single layer sheet and the twisting angle of the folded double-layer graphene. (b5) is the schematic image of the two folded double-layer parts shown in (b1). (c) shows the Raman spectra of the corresponding folded double-layer graphene sheets and together with the single layer graphene shown in (a1) and (b1). The excitation energy here is $E_{laser}$ = 2.33 eV.

**Figure 2.** Laser-excitation-energy dependent Raman spectra of (a) single layer graphene, (b) folded double-layer graphene of $\theta_{small}$, (c) folded double-layer graphene of $\theta_{medium}$, and (d) folded double-layer graphene of $\theta_{large}$.

**Figure 3**. (a) The frequency separation of the G'+ mode and G'- mode as a function of the laser-excitation-energy. Here, we fitted the G' mode of the1+1 $\theta_{medium}$ sample by two Lorentzian peaks. (b) and (c) are Raman spectra of G' mode with fitted curves of folded double-layer graphene of $\theta_{medium}$ done with different light polarization under the excitation energy of 2.54 eV.

**Figure 4**. Panels a1(a5) and b1(b5) show the optical images (schematic images) of different folded tetra-layer graphene samples. Panels a2(a6), a3(a7) and a4(a8) are Raman images of the G(G') mode integrated intensity, the G(G') mode frequency, and the G(G') mode width of the corresponding folded tetra-layer graphene samples shown in panel a1. Panels b2(b6), b3(b7) and b4(b8) are Raman images of the G(G') mode integrated intensity, the G(G') mode frequency, and the G(G') mode width of the corresponding folded tetra-layer graphene samples shown in panel b1. (c) shows the Raman spectra of the corresponding folded tetra-layer graphene sheets and together with the AB-stacked bilayer graphene shown in (a1) and (b1). The excitation energy here is $E_{laser} = 2.33$ eV.

**Figure 5**. (a-c) are laser-excitation-energy dependent Raman spectra of folded tetra-layer graphene of $\theta_{medium}$. (d) and (e) are the peak width of the G+ mode and G- mode and the frequency separation between the G+ mode and G- mode as a function of temperature under the excitation energy of 2.33 eV for folded tetra-layer graphene of $\theta_{medium}$, respectively. (f) Raman spectrum of G mode with fitted curves shown in panel b.

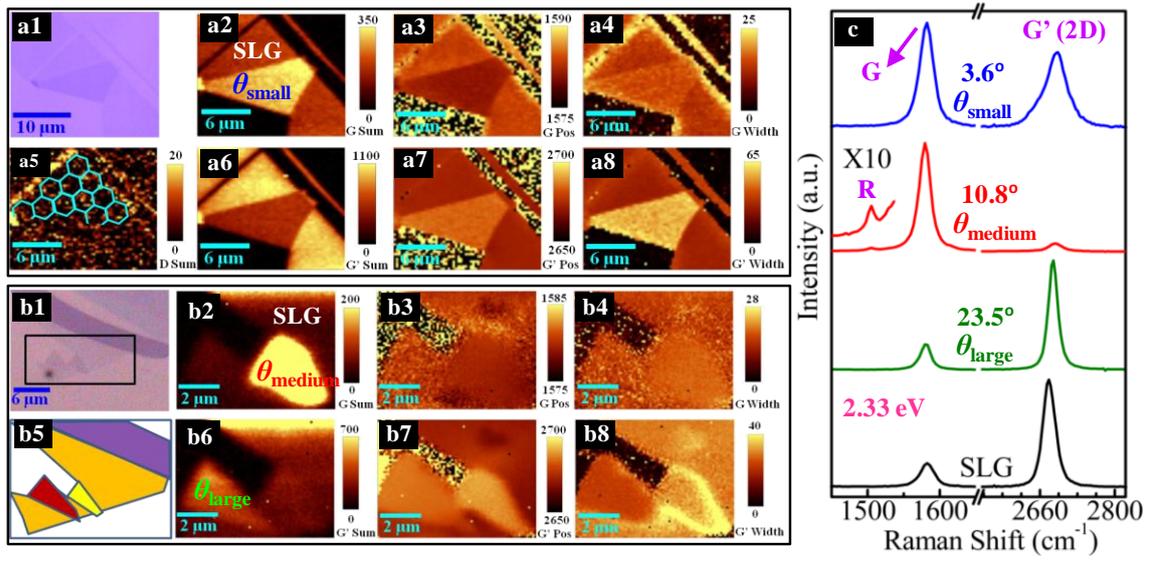

**Figure 1**

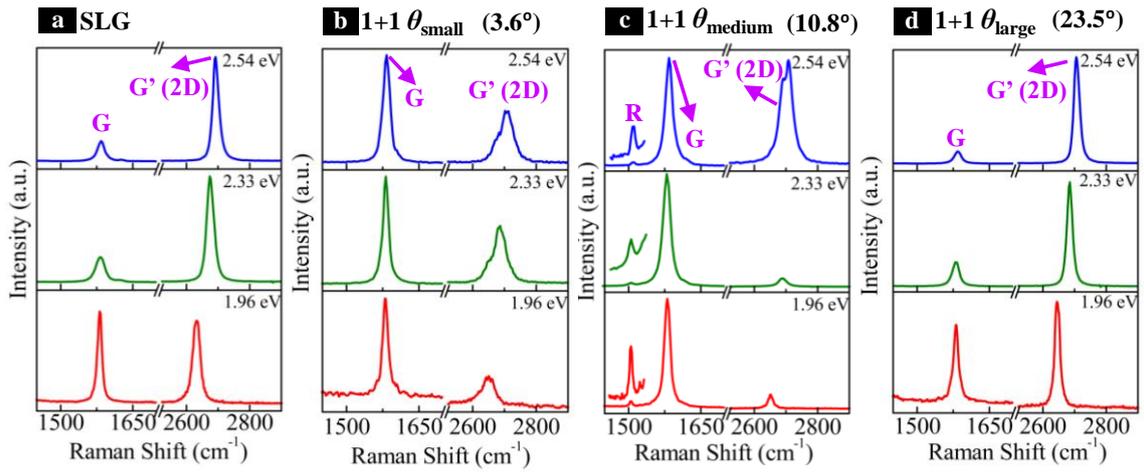

**Figure 2**

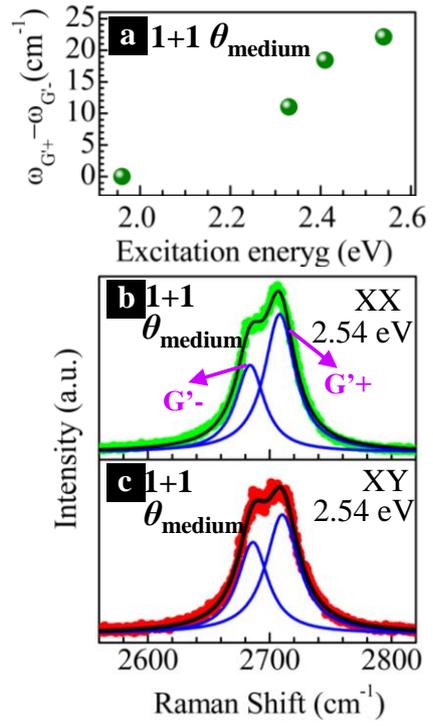

**Figure 3**

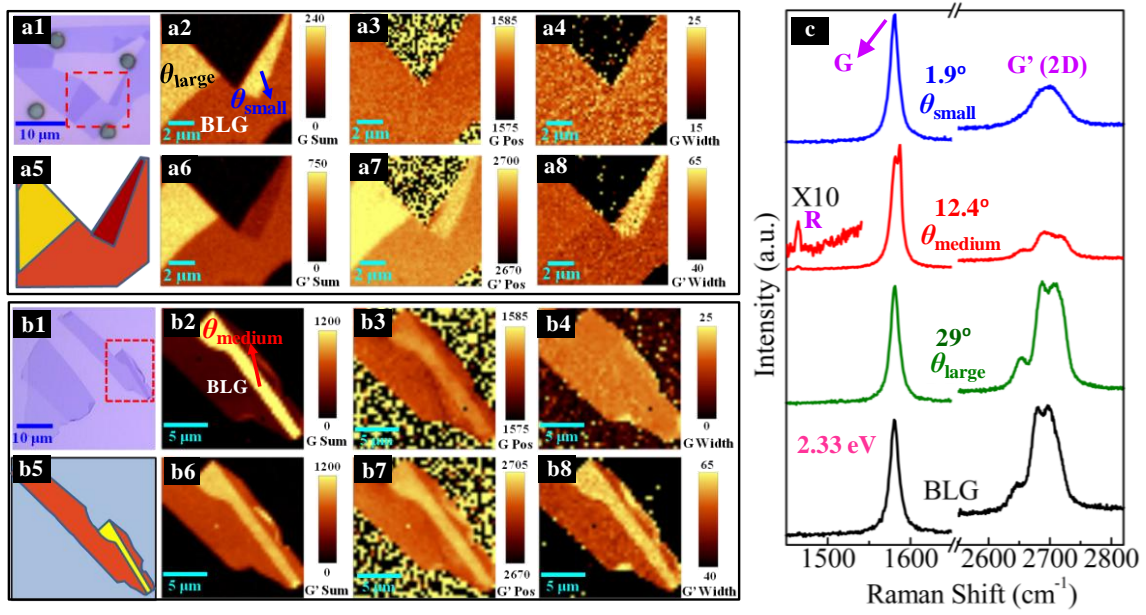

**Figure 4**

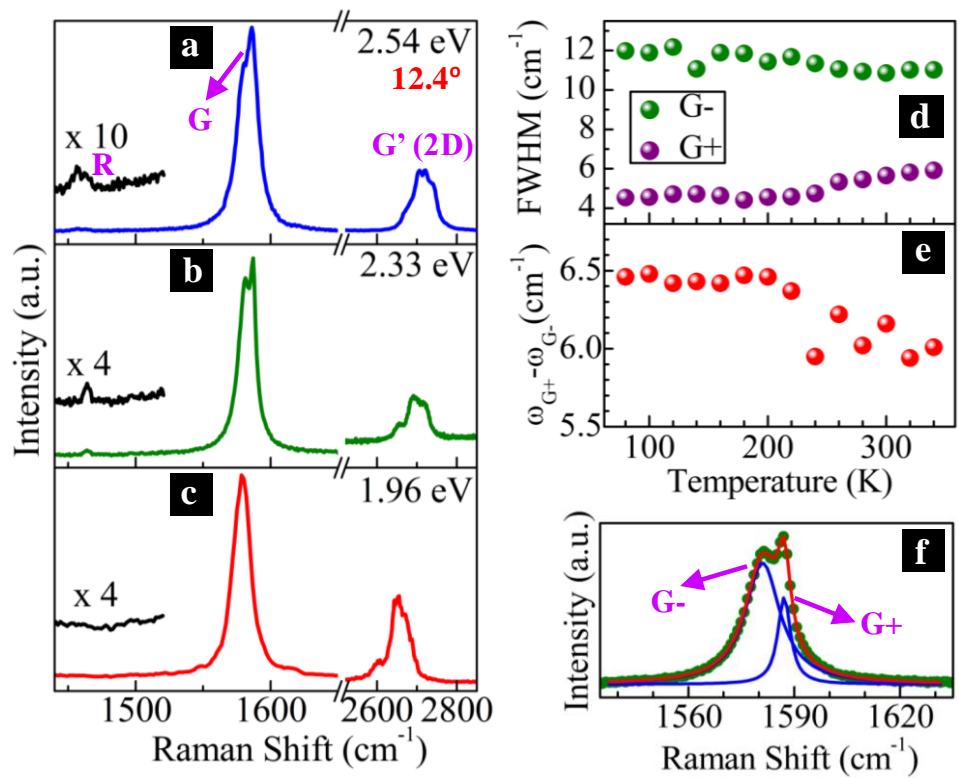

**Figure 5**

# Support information

# Evolution of Raman G and G' (2D) Modes in Folded Graphene Layers

Chunxiao Cong,[1] Ting Yu[1,2,3] [†]

[1]*Division of Physics and Applied Physics, School of Physical and Mathematical Sciences, Nanyang Technological University, 637371, Singapore;*
[2]*Department of Physics, Faculty of Science, National University of Singapore, 117542, Singapore;*
[3]*Graphene Research Center, National University of Singapore, 117546, Singapore*

[†]Address correspondence to yuting@ntu.edu.sg

There are five parts in this Supporting Information:

1. Rotational angles determination.

2. G' mode evolution of 1+1 $\theta_{small}$ folded double layer graphene together with that of AB-stacked BLG under different excitation photon energies.

3. Schematic diagram of outer and inner scattering processes of G' band in 1+1 fDLG of $\theta_{medium}$.

4. Doublet splitting of G mode in another 2+2 f4LG samples of $\theta_{medium}$.

5. In-situ temperature-dependent Raman spectra of 2+2 $\theta_{medium}$ folded tetra-layer graphene.

1. Rotational angle determination

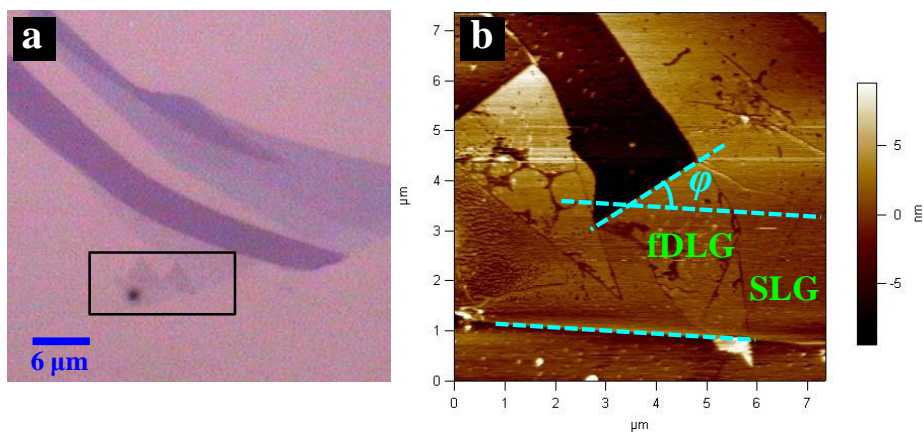

Figure S1. (a) Optical image of a folded double layer graphene (fDLG) sample on SiO$_2$/Si substrate that contains two folded parts. (b)Atomic force microscopy (AFM) image of the fDLG sample shown in (a). We only take one folded part as an example to show how to determine the twisting angle. The crystal axis of the single layer graphene (SLG) (i.e., the relative sharp and straight edge of theSLG) as well as folding line is shown by the aqua dashed line in the AFM image. The angle between the crystal axis and the folding line is marked by $\varphi$.

The crystal axisof SLG can be determined by the relative sharp and straight graphene edge. According to the geometry analysis, the rotation angle $\theta$, which is the angle of the top layer rotated relative tothe bottom layer, can be determined as $\theta = 2\varphi$ or $\theta = 180°-2\varphi$. In this work the angle $\varphi$ were measured 10 times to minimize the measurement error. Then we obtained 10 sets of rotation angle $\theta$ shown in the table below.

|   | 1 | 2 | 3 | 4 | 5 | 6 | 7 | 8 | 9 | 10 |
|---|---|---|---|---|---|---|---|---|---|----|
| $\theta$ | 10.32° | 11.78° | 10.28° | 10.2° | 11.06° | 10.64° | 11.56° | 10.94° | 10.76° | 10.46° |

The average value and the standard deviation of these 10 sets of rotation angle $\theta$ are 10.8° and 0.54, respectively.

2. G' mode evolution of 1+1 $\theta_{small}$ folded double layer graphene (fDLG)

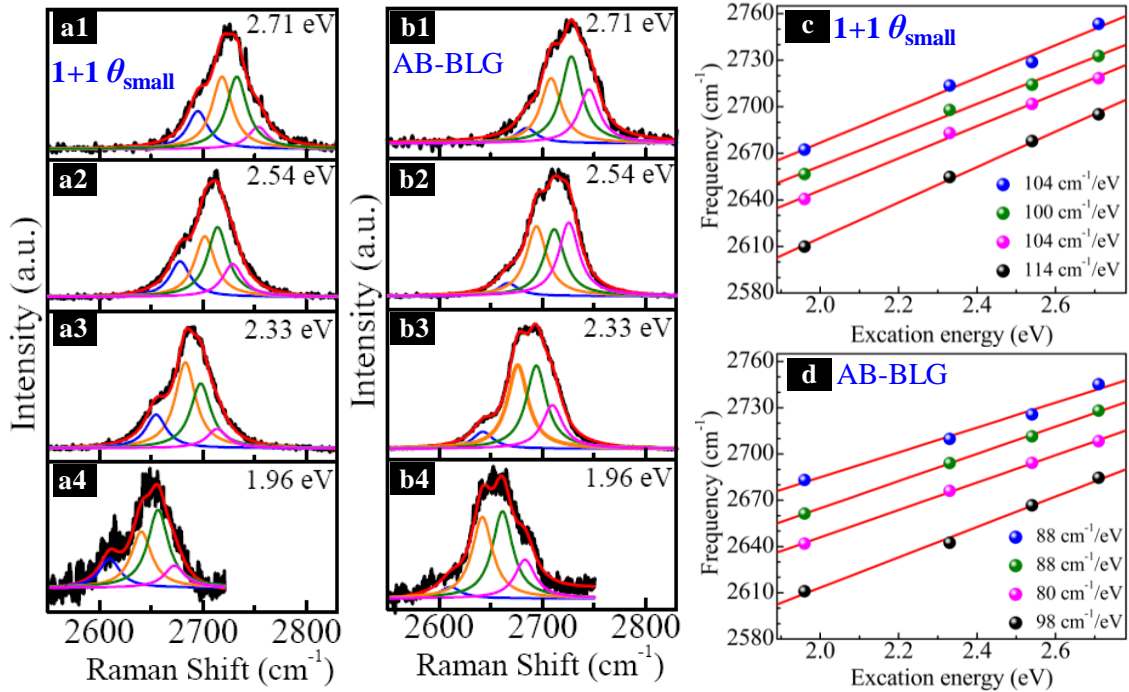

**Figure S2**. (a1-a4) and (b1-b4) are laser-excitation-energy dependent Raman spectra of G' band with fitted curves of folded double-layer graphene of $\theta_{small}$ and AB-stacked bilayer graphene, respectively. (c) and (d) are plots of the G' band frequencies of the folded double-layer graphene of $\theta_{small}$ and AB-stacked bilayer graphene as a function of excitation energy. We fitted the G' band of folded double-layer graphene of $\theta_{small}$ and AB-stacked bilayer graphene as four Lorentzian peaks. The dispersion behaviour for both samples is clearly seen.

3. Schematic diagrams of outer and inner scattering processes of G' band in 1+1 fDLG of $\theta_{medium}$.

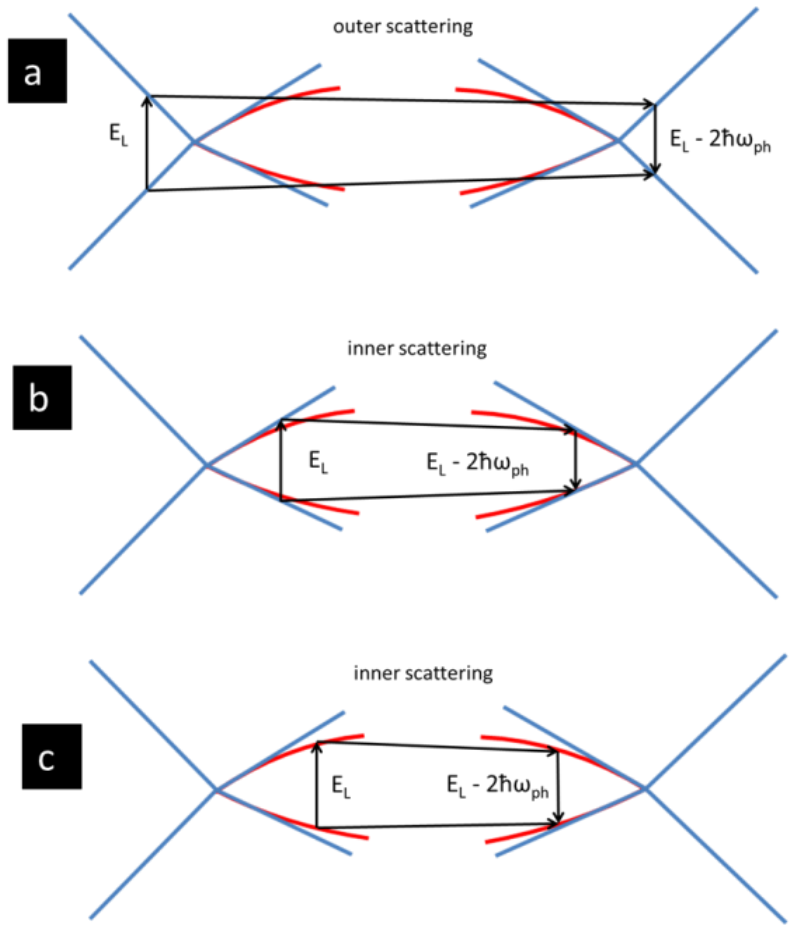

**Figure S3**. Schematic diagrams of outer and inner scattering processes in pristine single layer graphene (a and b) and $\theta_{medium}$ 1+1 fDLG (c). The curves in red represent the dispersion along *K-M* direction (inner scattering) of $\theta_{medium}$ 1+1 fDLG, which are further bent towards lower energy by the twist.[1]

4. Doublet splitting of G mode in another 2+2 f4LG sample of $\theta_{medium}$.

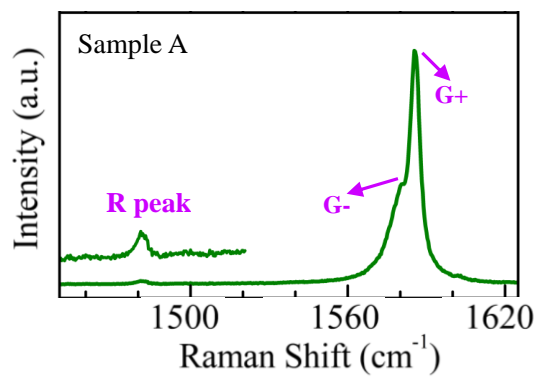

**Figure S4**. Raman spectrum of doublet feature of the G mode of $\theta_{medium}$ f4LG sample. The folding angle is 11.8 degrees.

**Table S1.** Fitted G peak parameters of $\theta_{medium}$ f4LG sample of Fig. S4 together with the fitted results of the sample in Fig. 5.

|  | Sample A | Sample of Fig. 5 |
| --- | --- | --- |
| $I_{G+}/I_{G-}$ | 1.25 | 0.31 |
| Width (G-)/Width (G+) | 2.33 | 2.31 |
| Frequency difference of G+ and G- | 6.3 | 6.2 |

5. *In-situ* temperature-dependent Raman spectra of G mode in 2+2 $\theta_{medium}$ folded tetra-layer graphene (f4LG)

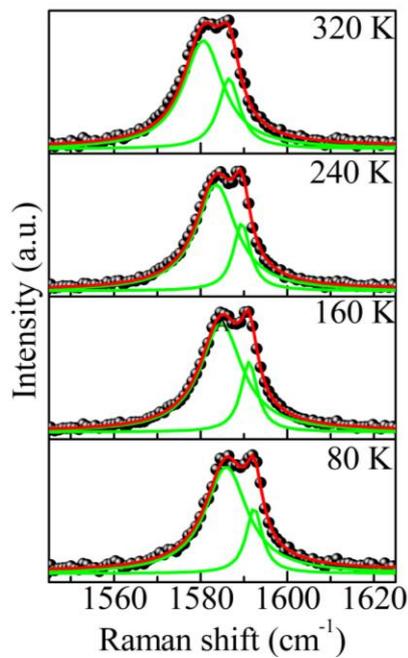

**Figure S5**. Temperature-dependent Raman spectra in the G mode region with fitted curves of folded tetra-layer graphene of $\theta_{medium}$.

**References:**

1. E. S. Morell, J. D. Correa, P. Vargas, M. Pacheco, Z. Barticevic, Phys. Rev. B **82**, 121407 (2010).